\documentclass[conference]{IEEEtran}
\IEEEoverridecommandlockouts
\usepackage{geometry}
\usepackage{cite}
\usepackage{amsmath,amssymb,amsfonts}
\usepackage{graphicx}
\usepackage{textcomp}
\usepackage{xcolor}
\usepackage{enumitem}
\usepackage{subcaption}
\usepackage{algorithm}
\usepackage{algorithmicx}
\usepackage[noend]{algpseudocode} 
\usepackage[export]{adjustbox}
\def\BibTeX{{\rm B\kern-.05em{\sc i\kern-.025em b}\kern-.08em
    T\kern-.1667em\lower.7ex\hbox{E}\kern-.125emX}}

\renewcommand{\baselinestretch}{0.96}

\begin{document}

\title{UAV-Assisted Coverage Hole Detection Using Reinforcement Learning in Urban Cellular Networks\\
\thanks{This research is supported in part by NSF grants CNS-1939334 and ECCS-2227499.}
}

\author{
\IEEEauthorblockN{Mushfiqur Rahman$^{\ast}$, \.{I}smail G\"{u}ven\c{c}$^{\ast}$, David Ramirez$^{\dagger}$, and Chau-Wai Wong$^{\ast}$}
\IEEEauthorblockA{
    $^{\ast}$Department of Electrical and Computer Engineering, North Carolina State University, Raleigh, NC, USA \\
    $^{\dagger}$DOCOMO Innovations, Inc., Sunnyvale, CA, USA \\
    Emails: \{mrahman7, iguvenc, chauwai.wong\}@ncsu.edu\\
    david.ramirez@docomoinnovations.com
}
}

\maketitle

\begin{abstract}
Deployment of cellular networks in urban areas requires addressing various challenges. For example, high-rise buildings with varying geometrical shapes and heights contribute to signal attenuation, reflection, diffraction, and scattering effects. This creates a high possibility of coverage holes (CHs) within the proximity of the buildings. Detecting these CHs is critical for network operators to ensure quality of service, as customers in these areas may experience weak or no signal reception.
To address this challenge, we propose an approach using an autonomous vehicle, such as an unmanned aerial vehicle (UAV), to detect CHs, for minimizing drive test efforts and reducing human labor. The UAV leverages reinforcement learning (RL) to find CHs using stored local building maps, its current location, and measured signal strengths. As the UAV moves, it dynamically updates its knowledge of the signal environment and its direction to a nearby CH while avoiding collisions with buildings.
We created a wide range of testing scenarios using building maps from OpenStreetMap and signal strength data generated by NVIDIA Sionna raytracing simulations. The results show that the RL-based approach outperforms non-machine learning, geometry-based methods in detecting CHs in urban areas. Additionally, even with a limited number of UAV measurements, the method achieves performance close to theoretical upper bounds that assume complete knowledge of all signal strengths.
\end{abstract}


\section{Introduction}
The advancements in Long-Term Evolution (LTE) and 5G technology have brought the hope of ubiquitous high data rates outdoors, enabling data-intensive and real-time services. To maintain quality of service (QoS), network operators must ensure that several key performance indicators (KPIs), such as latency and throughput, are met regardless of a user's location. However, weak signal reception can cause the wireless network to underperform in certain areas, failing to meet QoS expectations. This presents a challenge for network operators, as they aim to provide consistent and reliable communication across a region by strategically deploying base stations~(BSs). The evolving urban landscape includes buildings and structures that complicate this effort, as they can create small zones with weak or no coverage, commonly referred to as coverage holes (CHs). These CHs are a significant concern for network operators, as they impact overall service quality and necessitate an effective method for coverage hole detection~(CHD).
\begin{figure}[!t]
    \centering

    \begin{subfigure}[b]{0.8\linewidth}
        \centering
        \includegraphics[width=\textwidth,trim={0.1cm 0.1cm 1.25cm 0.1cm},clip]{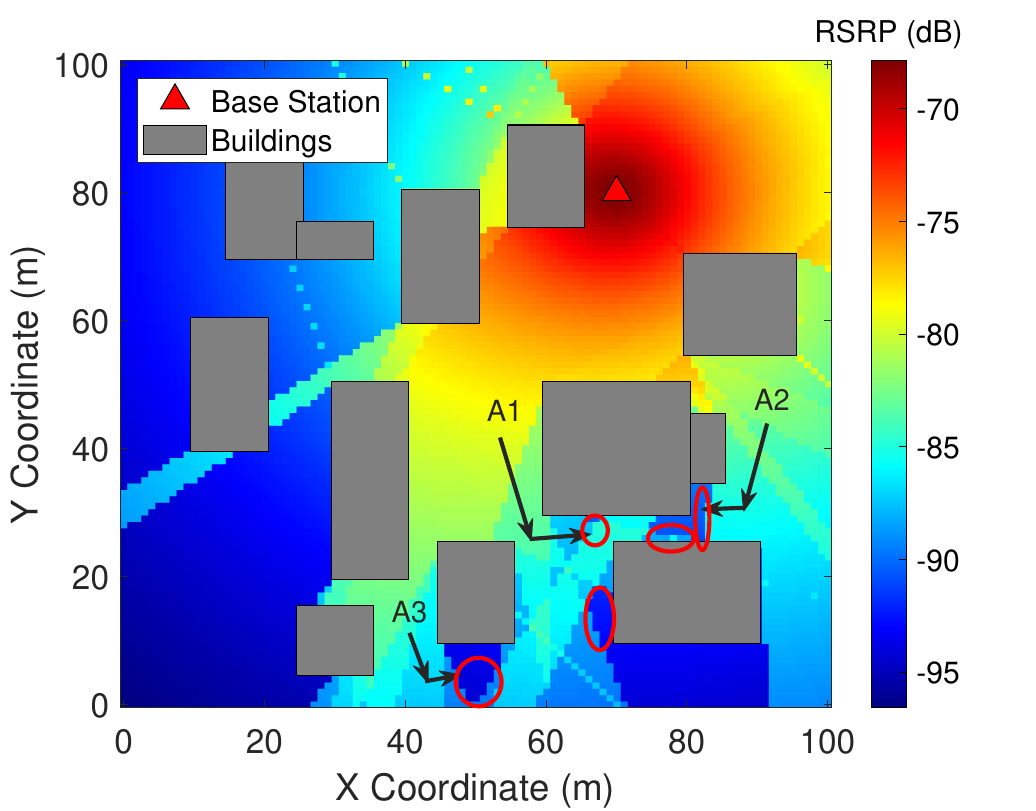}
    \end{subfigure}

    \caption{The goal of this work is to deploy a UAV to autonomously detect coverage holes in an urban area using building maps. The challenge is to guide the UAV from an initial location to a nearby coverage hole (examples circled in red) while avoiding collisions with buildings.}
    \vspace{-0.5cm}
    \label{fig:example}
\end{figure}

The network operators traditionally rely on expensive drive tests (DTs) involving manual efforts and propagation models, such as ray tracing, to construct a coverage map (CM) for CHD in an area~\cite{liang2016coverage}. However, these CMs, derived from sparse drive test data and simulation models, often produce inaccurate results~\cite{al2020optimal}. Additionally, drive tests are costly and must be performed periodically to account for construction changes, environmental factors, and seasonal variations, such as changes in tree coverage and weather conditions. This requires network operators to continuously identify and address coverage holes to maintain QoS with added operational expenditure (OPEX) even after the initial deployment of BSs. For example, in Germany, LTE deployments have uncovered coverage holes, which have been marked in maps~\cite{al2020optimal}. In Scotland, the government, in collaboration with the Scottish Futures Trust, initiated the Scottish 4G Infill Programme~\cite{scottish2018} to address approximately $50$–$60$ coverage holes in LTE networks. Similarly, in Japan, network operators are actively working to cover the CHs~\cite{ebuchi2019kddi}. These initiatives demonstrate the importance of addressing CHs to ensure reliable network performance.

To reduce the cost of DT, the minimization of drive test~(MDT) feature was introduced, starting with the 3rd Generation Partnership Project (3GPP) standards for 4G networks. MDT enables the collection of Radio Link Failure (RLF) reports from user equipment (UE) as an alternative to the DT. These RLF reports contain the UE's location along with signal strength measurements recorded whenever a connection is lost, and the report is sent to the target BS or the next available BS in the handover process~\cite{3gpp2015}. However, MDT is not without limitations. RLF reports are only sent when the UE reconnects to the serving BS or successfully hands over to another BS, which may not occur if a large CH exists. Additionally, the geographic location of the UE may be inaccurate or outdated, depending on the user's location settings. Privacy concerns also arise, as customers may restrict the collection of RLF reports under regulations such as the General Data Protection Regulation (GDPR)\cite{al2020optimal}. Furthermore, constructing a radio map from RLF data requires propagation models or interpolation mechanisms, which can introduce additional errors\cite{liang2016coverage}.

In light of these challenges, recent studies have proposed the use of autonomous vehicles, such as unmanned aerial vehicles~(UAVs) and unmanned ground vehicles (UGVs), as a cost-effective solution for CHD~\cite{al2020optimal, fan2024uav} and CM construction~\cite{zeng2021simultaneous, yang2024channel}. Specifically, we draw inspiration from the proposal by Al-Ahmed~\textit{et al.}~\cite{al2020optimal} to utilize reinforcement learning~(RL) to enable a UAV to autonomously detect CHs. In this work, we explore the feasibility of applying RL in urban environments with buildings through detailed simulation. The main contributions of this paper are summarized as follows:
\begin{itemize}
\item We demonstrate, through thorough simulations of building environments, that RL-guided UAVs can detect CHs with greater precision and recall compared to non-ML methods based on building geometry information.
\item Our experimental results, using signal strength data generated with NVIDIA Sionna, reveal that coverage holes are disproportionately more frequent near buildings. 
\item We show that, while a gradient-based method can guide UAVs toward CHs using known signal strength, RL can achieve similar results with far fewer measurements.
\end{itemize}


\section{Related Works}
\label{sec:rel_works}
The use of RL-based UAV trajectory planning in urban environments with buildings is a highly active research area in wireless communication. 
These UAV-based methods propose UAV trajectories within urban environments to achieve specific goals while ensuring aerial safety.
Most existing studies focus either on optimizing UAV trajectories to avoid obstacles including buildings, with objectives such as minimizing time, distance, or energy~\cite{guerra2023reinforcement,chen2022environment,dabiri2024optimizing}, or maintaining reliable communication with BS avoiding CHs~\cite{wang2021learning, yu2022uav, im2023trajectory, zhong2024joint, esrafilian20203d}. Illustrations of these tasks are provided in Fig.~\ref{fig:example}(a) and (b).
\begin{figure}[!t]
    \centering
    \begin{subfigure}[b]{0.49\linewidth}
        \centering
    \includegraphics[width=\textwidth,trim={0.1cm 0.1cm 1.05cm 0.1cm},clip]{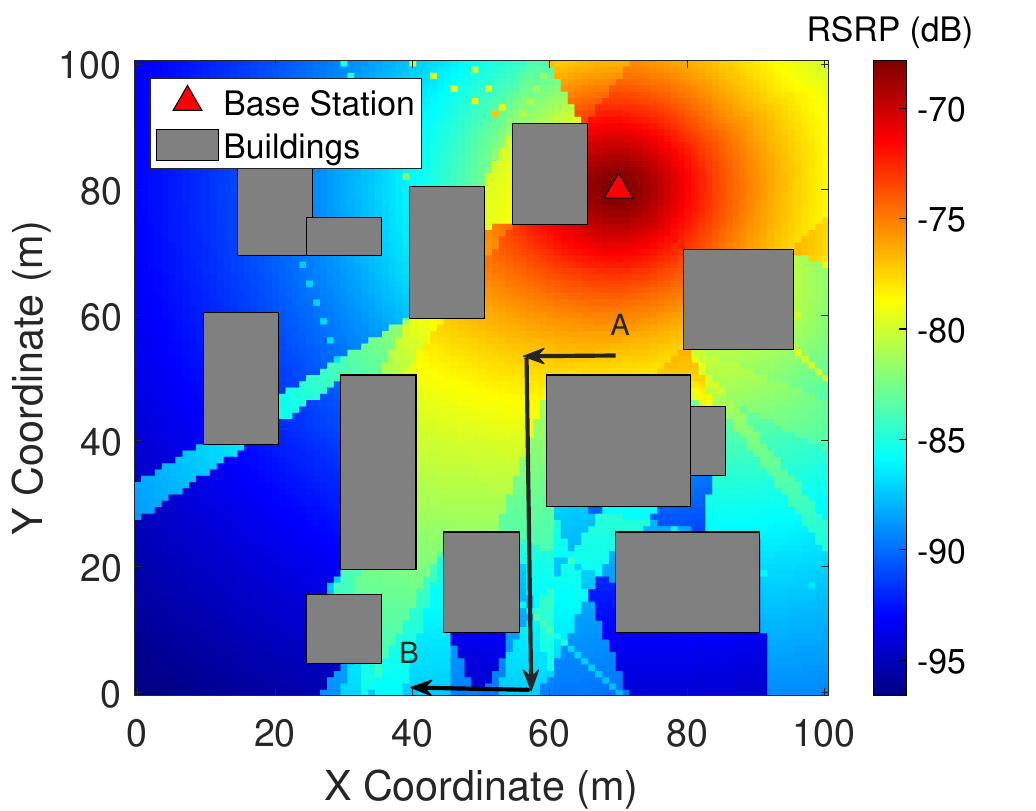}
        \caption{Obstacle avoidance}
    \end{subfigure}
    \hfill
    \begin{subfigure}[b]{0.49\linewidth}
        \centering
        \includegraphics[width=\textwidth,trim={0.1cm 0.1cm 1.05cm 0.1cm},clip]{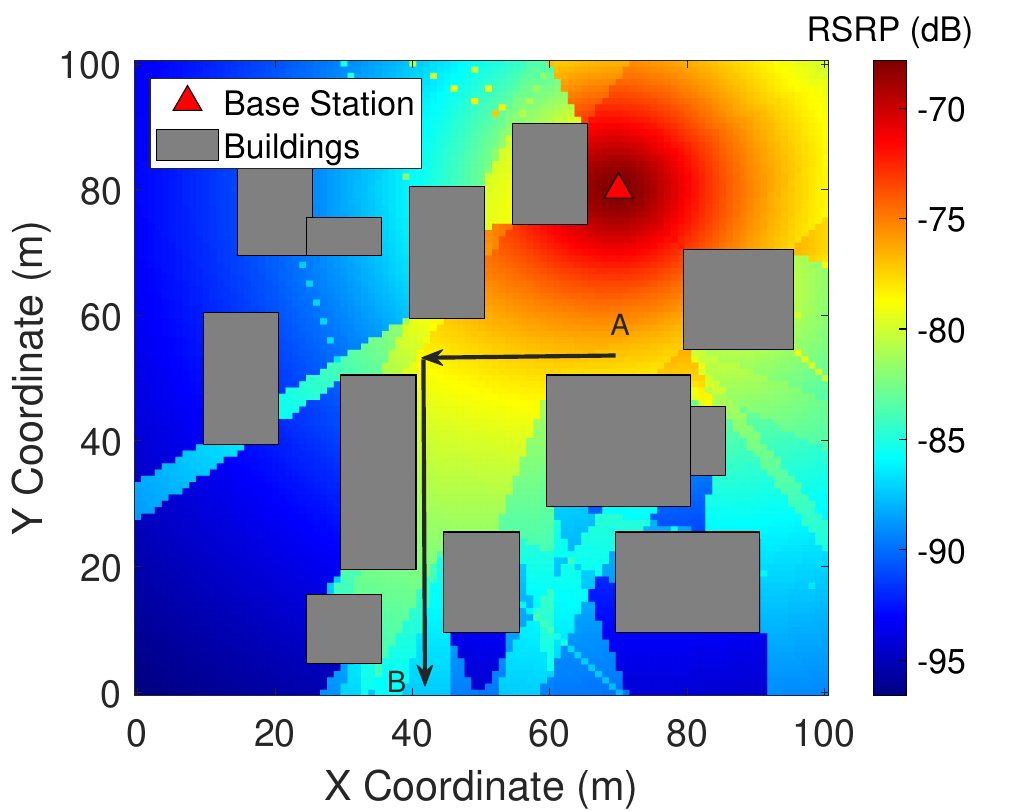}
        \caption{Coverage hole avoidance}
    \end{subfigure}

    \caption{Recent works on RL-enabled UAVs in the context of urban cellular networks with building infrastructure include: (a) UAV avoiding buildings with the goal of reaching a target location from a starting location, and (b) UAV avoiding weak signal regions to reach the target location from a starting~location.}
    \vspace{-0.5cm}
    \label{fig:example2}
\end{figure}
Some studies also explore multi-task learning approaches that aim to maintain reliable communication while constructing radio maps, possibly avoiding redundant locations to cover larger areas~\cite{zeng2021simultaneous, yang2024channel, hao2023deep, huang2021simultaneous, romero2020aerial, zhao2023online, gesbert2022uav}. However, these studies do not explicitly address CH detection. Other research in UAV-based BSs (UAV-BS) focuses on deploying UAVs as aerial BSs to serve more users or to temporarily cover CHs, maintaining a backhaul link to a ground BS~\cite{dabiri2023optimal, ma2023time, hu2022deep, fu2023joint, bayerlein2018trajectory, bayerlein2018learning, challita2019interference}. These RL-based methods use reward functions to optimize UAV trajectories, locations, or orientations to maximize aggregated throughput or serve the maximum number of ground UEs. Although these reward functions favor positioning UAVs near CHs, the algorithms do not directly identify the locations or sizes of CHs, which is essential for adjusting BS configurations to address them without deploying UAV-BSs.

To our knowledge, only Fan~\textit{et al.}~\cite{fan2024uav} and Al-Ahmed~\textit{et al.}~\cite{al2020optimal} have investigated UAV-assisted CH detection. Fan ~\textit{et al.}\cite{fan2024uav} rely on a preset trajectory based on known BS locations, traversing boundary areas between BSs to identify potential CHs without using machine learning (ML). In contrast, Al-Ahmed~\textit{et al.}\cite{al2020optimal} employ RL to guide UAVs for CH detection but do not consider buildings or obstacles in their simulations. To the best of our knowledge, this is the first paper to present results on CH detection in an environment with building structures, reflecting typical urban scenarios.

\section{System Model And Problem Formulation}
\label{sec:sys_model}
We consider a UAV performing as an RL agent for CH detection to have offline access to the building map of its navigating area. The UAV follows a dynamic trajectory, with each next waypoint predicted by the RL algorithm in real time based on its current and previous waypoints. At each waypoint, it measures the reference signal received power (RSRP) from the desired BS, stores the value, and uses the RL algorithm to determine the next waypoint. The ultimate goal of the UAV is to locate a CH, where the RSRP falls below a predefined threshold, $\epsilon_{\text{CH}}$. The UAV operates at a fixed altitude $H$ above the ground, limiting its movement to a horizontal plane for traversal and coverage hole discovery.

\begin{figure}[!t] 
    \centerline{
    \includegraphics[width=\linewidth,trim={0.6cm 18.7cm 2.3cm 0cm},clip]{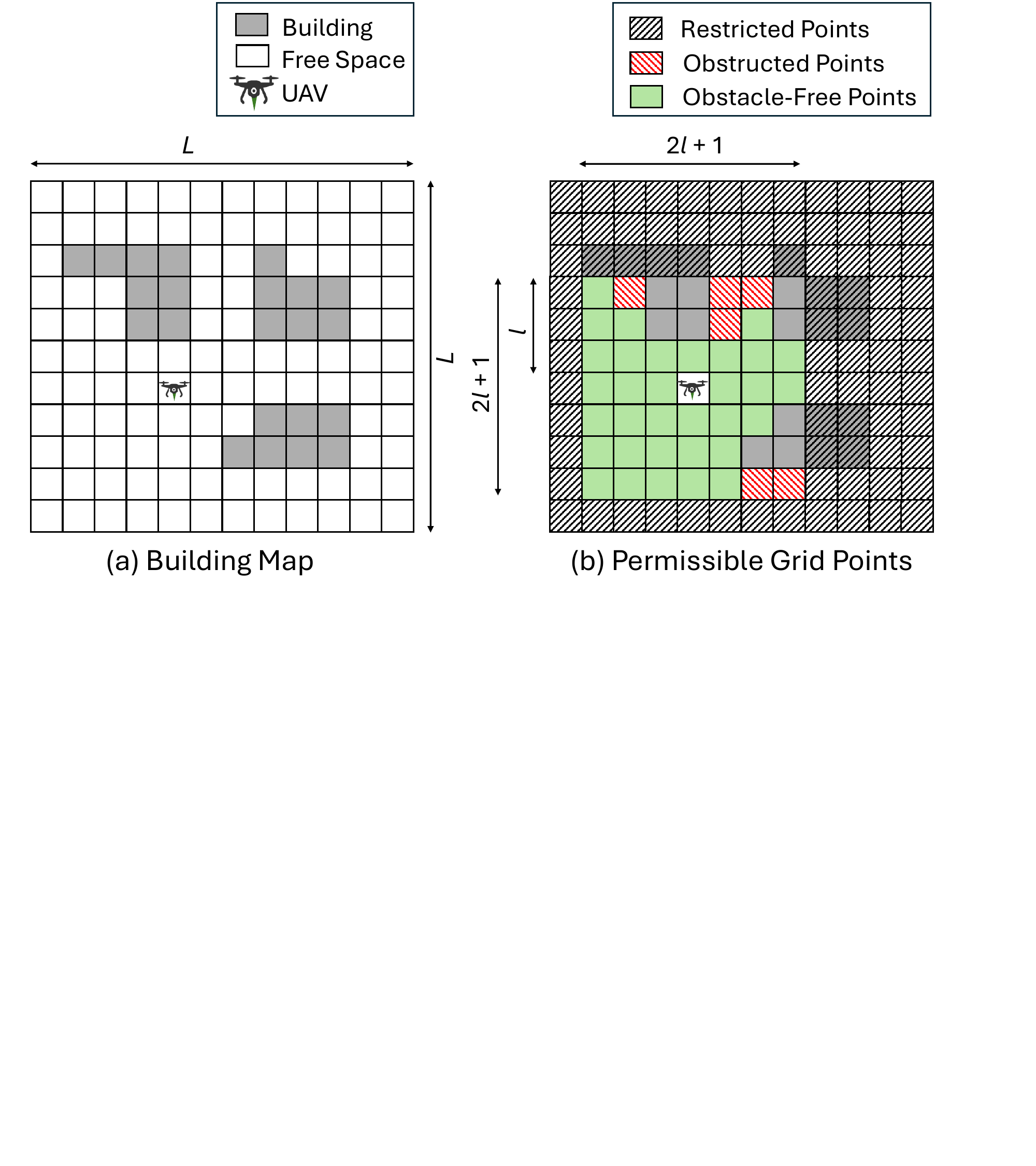}} 
    \caption{The given building map is divided into square grids, with the UAV positioned at one of the grid points. At each iteration, the UAV can move to another grid point, subject to the following conditions: (a) it cannot move through building obstructions, and (b) in a single step, it can move up to a maximum of $l$ grid points in the left, right, up, or down directions. The green-marked grid points within the \((2l+1) \times (2l+1)\) matrix represent the valid locations the UAV can move to in the next iteration, excluding grid points where buildings obstruct the straight-line path. The figure is best viewed in color.}
    \vspace{-0.5cm}
    \label{fig:system_model} 
\end{figure}

Let us define the building map known to the UAV as being divided into an $L \times L$ grid, where each grid point is represented as $\mathbf{S}(i,j)$, with $i \in \{1,2,\dots,L\}$ and $j \in \{1,2,\dots,L\}$. Let the building map be denoted as $\mathbf{M}_{\text{bld}} \in \mathbb{R}^{L \times L}$, where each element represents the building height in meters, with a value of zero indicating the absence of a building. It can be noted that the square building map is solely for simplicity, without losing the generality of its shape. The binary building occupancy map at the fixed altitude $H$ is illustrated in Fig.~\ref{fig:system_model}(a). 
Let us denote the set of grid points occupied by buildings as $\mathbf{S}_{\text{bld}}$, defined as:
\begin{equation}
\mathbf{S}_\text{bld} = \{\mathbf{S}(i, j) \in (\mathbb{N}, \mathbb{N}) \mid \mathbf{M}_\text{bld}(i, j) \geq H\}.
\end{equation}
Let the grid location of the UAV at the \( k \)-th iteration be denoted by \( \mathbf{S}^{\text{UAV}}(k) \in (\mathbb{N},\mathbb{N}) \), and the measured RSRP at this location be \( Z(k) \in \mathbb{R}\). As illustrated in Fig.~\ref{fig:example}, the goal of the UAV is to reach a CH in as few iterations as possible, starting from an initial location. For each starting point A1, A2, or A3, as illustrated in Fig.~\ref{fig:example}, the UAV needed \( 2 \) iterations to reach~a~CH.

At the \( k \)-th iteration, the UAV has collected \( k \) measurements from its visited locations \( \mathbf{S}^{\text{UAV}}(i) \), where \( i \in \{1,2,\dots,k\} \). Let the set of these measurements and locations be denoted as \( \mathbf{Z}_{\text{cmb}}(k) \) and \( \mathbf{S}^{\text{UAV}}_{\text{cmb}}(k) \), respectively:
\begin{equation}
\mathbf{Z}_{\text{cmb}}(k) = \bigcup_{i=1}^{k} Z(i), \quad
\mathbf{S}^{\text{UAV}}_{\text{cmb}}(k) = \bigcup_{i=1}^{k} \mathbf{S}^{\text{UAV}}(i).
\end{equation}
When the UAV chooses its next waypoint \( \mathbf{S}^{\text{UAV}}(k+1)\), it must ensure that the straight-line path from the current waypoint to the next waypoint is not obstructed by any of the buildings. Additionally, the UAV is allowed to move a maximum of $l$ grid points in each of the four cardinal directions, forming a \((2l+1) \times (2l+1)\) square matrix. Considering the maximum displacement at each iteration and the building occupancy constraints, the valid grid points for the next waypoint are shown as solid green points in Fig.~\ref{fig:system_model}(b).

Considering the current UAV location, \( \mathbf{S}^{\text{UAV}}(k)\), let us define $\mathbf{S}_{\text{blk}}(k)$ as the grid points where the straight-line paths from \( \mathbf{S}^{\text{UAV}}(k)\) are blocked by buildings, and $\mathbf{S}_{\text{alw}}(k)$ as the grid points within a maximum horizontal displacement of $l$ from \( \mathbf{S}^{\text{UAV}}(k)\) in any cardinal direction. The set of permissible grid points $\mathbf{S}_{\text{prm}}(k)$, which the UAV can select as its next waypoint, is then expressed as:
\begin{equation}
\mathbf{S}_{\text{prm}}(k) = \mathbf{S}_{\text{alw}}(k) \setminus (\mathbf{S}_{\text{bld}} \cup \mathbf{S}_{\text{blk}}(k)).
\label{eqn:S_prm}
\end{equation}
The objective is to learn a function \( \phi \) that predicts the next waypoint \( \mathbf{S}^{\text{UAV}}(k+1) \in \mathbf{S}_{\text{prm}}(k) \) based on the combined locations \( \mathbf{S}^{\text{UAV}}_{\text{cmb}}(k) \), the combined measurements \( \mathbf{Z}_{\text{cmb}}(k) \), and the building map \( \mathbf{M}_{\text{bld}} \):
\begin{equation}
\mathbf{S}^{\text{UAV}}(k+1) = \phi \Big(\mathbf{S}^{\text{UAV}}_{\text{cmb}}(k), \mathbf{Z}_{\text{cmb}}(k), \mathbf{M}_{\text{bld}} \Big),
\end{equation}
with the goal of minimizing the number of iterations required to reach the CH. The number of iterations required to reach the CH can be expressed using a recursive equation. The optimization problem is defined as:
\begin{equation}
\min_{\phi} N_\phi(k) \quad \text{subject to } \mathbf{S}^{\text{UAV}}(k+1) \in \mathbf{S}_{\text{prm}}(k),
\end{equation}
where \( N_\phi(k) \), the number of iterations to reach the CH, is given by:
\begin{equation}
N_\phi(k) =
\begin{cases} 
    0 & \text{if } Z(k) \leq \epsilon_{\text{CH}}, \\
    1 + N_\phi(k+1) & \text{otherwise}.
\end{cases}
\end{equation}
This formulation ensures that the UAV minimizes its iterations for finding CH with \(Z(k) \leq \epsilon_{\text{CH}}\) while adhering to the constraints of valid movement defined by \( \mathbf{S}_{\text{prm}}(k) \).

\section{DDQN-Based CH Detection}
\label{sec:pro_method}
We employ a Double Deep Q-Network (DDQN)~\cite{van2016deep} for the autonomous detection of coverage holes, utilizing the UAV as an RL agent.
DDQN is an enhancement of the Deep Q-Network (DQN) that addresses the overestimation issue of Q-values by using two networks: the policy network and the target network.
The key design features of our DDQN model, along with its application in UAV trajectory updates, are outlined below.

\subsection{State and Action Space}
The UAV searching for CHs predicts waypoints sequentially at each step to dynamically update its trajectory. As discussed in Section~\ref{sec:sys_model}, the UAV knows \( k \) RSRP measurements at the \( k \)-th iteration and has access to an offline building height map of the area. These combined features constitute the state of the UAV in the RL framework.

\begin{figure}[t!]
    \centering
    \begin{subfigure}[b]{0.49\linewidth}
        \centering
    \includegraphics[width=\textwidth,trim={0.1cm 0.1cm 0.0cm 0.1cm},clip]{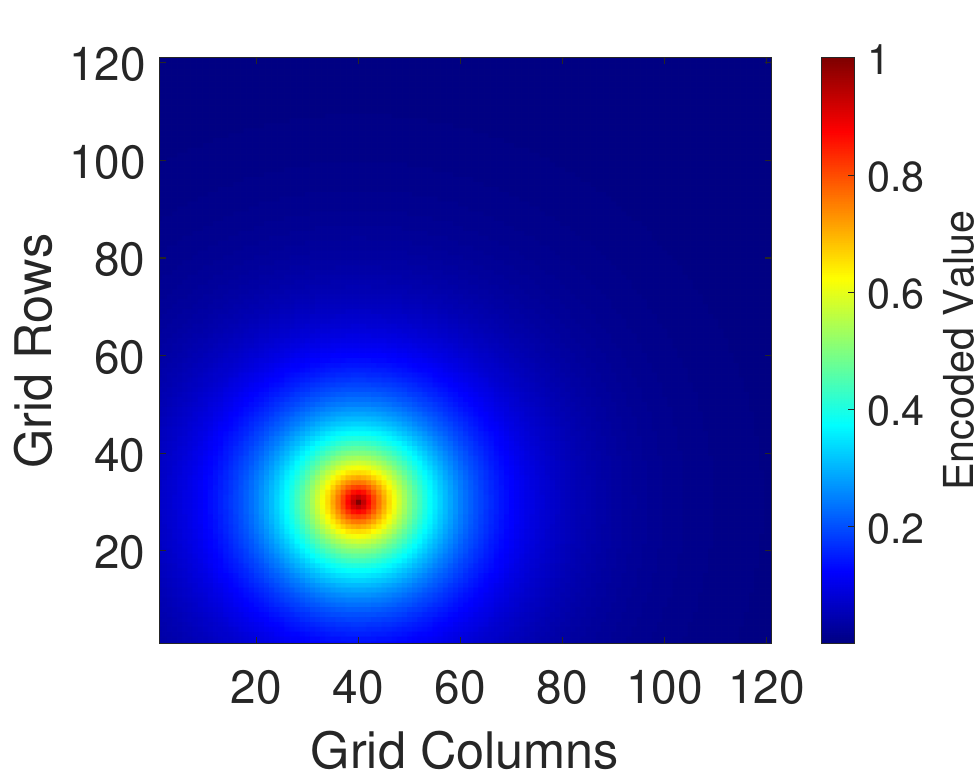}
        \caption{Location Encoding}
    \end{subfigure}
    \hfill
    \begin{subfigure}[b]{0.49\linewidth}
        \centering
        \includegraphics[width=\textwidth,trim={0.1cm 0.1cm 0.0cm 0.1cm},clip]{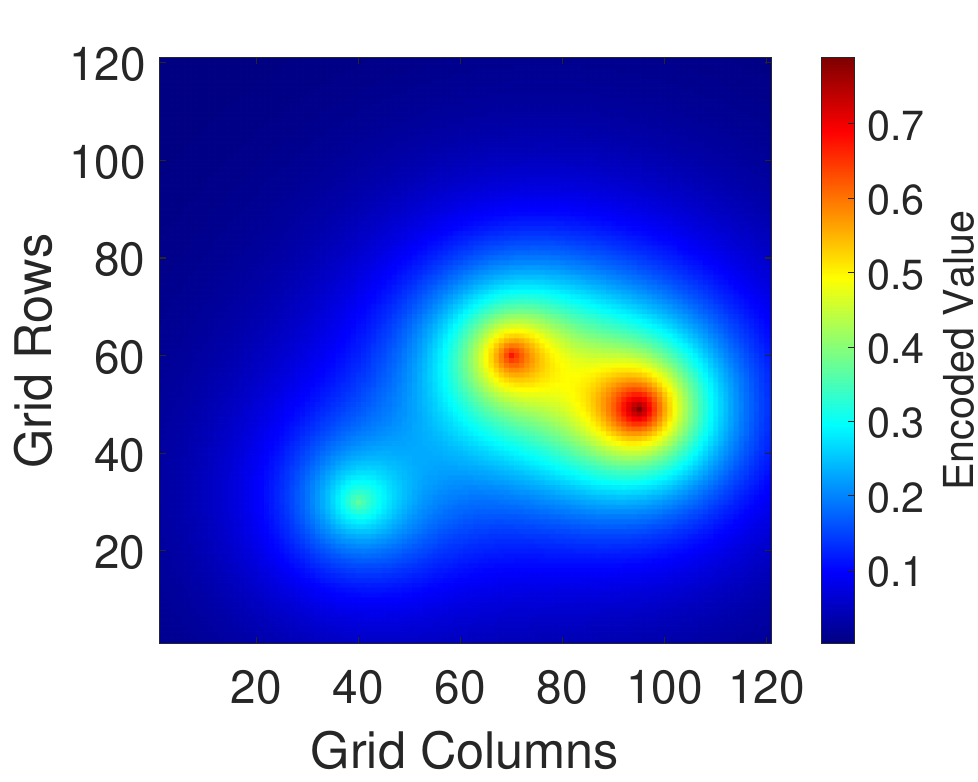}
        \caption{RSRP Encoding}
    \end{subfigure}

    \caption{Generated \(L \times L\) matrices by encoding (a) the current UAV location and (b) the current and past RSRP measurements for iteration number \(k=3\).}
    \vspace{-0.5cm}
    \label{fig:encoded_values}
\end{figure}

The grid location at \( k \)-th iteration of the UAV, \( \mathbf{S}^{\text{UAV}}(k) \), is encoded as a 2D array denoted by \( \mathbf{S}_{\text{enc}}^{\text{UAV}}(k) \in \mathbb{R}^{L \times L} \). The encoding is designed as a circular gradient centered at the UAV's location, where the value decreases exponentially with the Euclidean distance from the UAV's grid position. Mathematically, the encoded grid location is expressed as:
\begin{equation}
\mathbf{S}_{\text{enc}}^{\text{UAV}}(k)(i,j) = 2^{-c\|\mathbf{S}(i,j) - \mathbf{S}^{\text{UAV}}(k)\|},
\label{eqn:encode_loc}
\end{equation}
where \( i = 1, 2, \dots, L \), \( j = 1, 2, \dots, L \), \( \mathbf{S}(i,j) \) represents the grid location, \(c\) is a constant, and \( \| \cdot \| \) denotes the Euclidean distance. A demonstration of encoded grid location is given in Fig.~\ref{fig:encoded_values}(a).
Similarly, we encode the \( k \) measurements along with their respective measurement locations as a 2D array, by additively combining \( k \) circular gradients centered at the measurement locations. The maximum value of each gradient is derived from the measured RSRP. First, we normalize the measured RSRP (in dB) as follows:
\begin{equation}
Z_{\text{nrm}}(k) = \frac{Z(k) - \epsilon_{\text{CH}}}{-\epsilon_{\text{CH}}}.
\end{equation}
This normalization ensures that the output is positive for any RSRP value greater than the coverage hole threshold \( \epsilon_{\text{CH}} \). Next, we encode the measurements up to \(k\)-th iteration as a 2D array, denoted by \(\mathbf{Z}_{\text{enc}}(k) \in \mathbb{R}^{L \times L}\), as follows:
\begin{equation}
\mathbf{Z}_{\text{enc}}(k)(i,j) = \sum_{n=1}^{k} Z_{\text{nrm}}(n) \cdot 2^{-c\|\mathbf{S}(i,j) - \mathbf{S}^{\text{UAV}}(n)\|}.
\label{eqn:encode_rsrp}
\end{equation}
 Fig.~\ref{fig:encoded_values}(b) illustrates an example of the encoded RSRP measurements after the UAV has visited three locations.

Given the encoded grid location and measurements at the \(k\)-th iteration, \(\mathbf{S}_{\text{enc}}^{\text{UAV}}(k)\) and \(\mathbf{Z}_{\text{enc}}(k)\), the state of the RL agent \(\mathbf{h}(k)\) is defined as follows:
\begin{equation}
\mathbf{h}(k) = \big(\mathbf{S}_{\text{enc}}^{\text{UAV}}(k), \mathbf{Z}_{\text{enc}}(k), \mathbf{M}_{\text{bld}}\big).
\label{eqn:current_state}
\end{equation}
The action space consists of the possible set of actions the RL agent can take. As described in Section~\ref{sec:sys_model}, the formulated problem allows a maximum grid movement of \(l\) in any cardinal direction. Consequently, the UAV can move within a \((2l+1) \times (2l+1)\) grid, as illustrated in Fig.~\ref{fig:system_model}(b). Let the grid location of the UAV at the \(k\)-th iteration, \(\mathbf{S}^{\text{UAV}}(k)\), be denoted as \((i_k, j_k)\). The action space is defined as a cropped grid around the UAV's current location as follows:
\begin{equation}
\mathbf{\mathcal{A}}(k) = \big\{\mathbf{S}(i, j) \bigm| |i - i_k| \leq l, \, |j - j_k| \leq l \big\}.
\label{eqn:Ak_eqn}
\end{equation}

\subsection{Q-network Architecture}
\begin{figure}[t!]
    \centering
    \includegraphics[width=\linewidth,trim={1.8cm 7.4cm 4.4cm 0.2cm},clip]{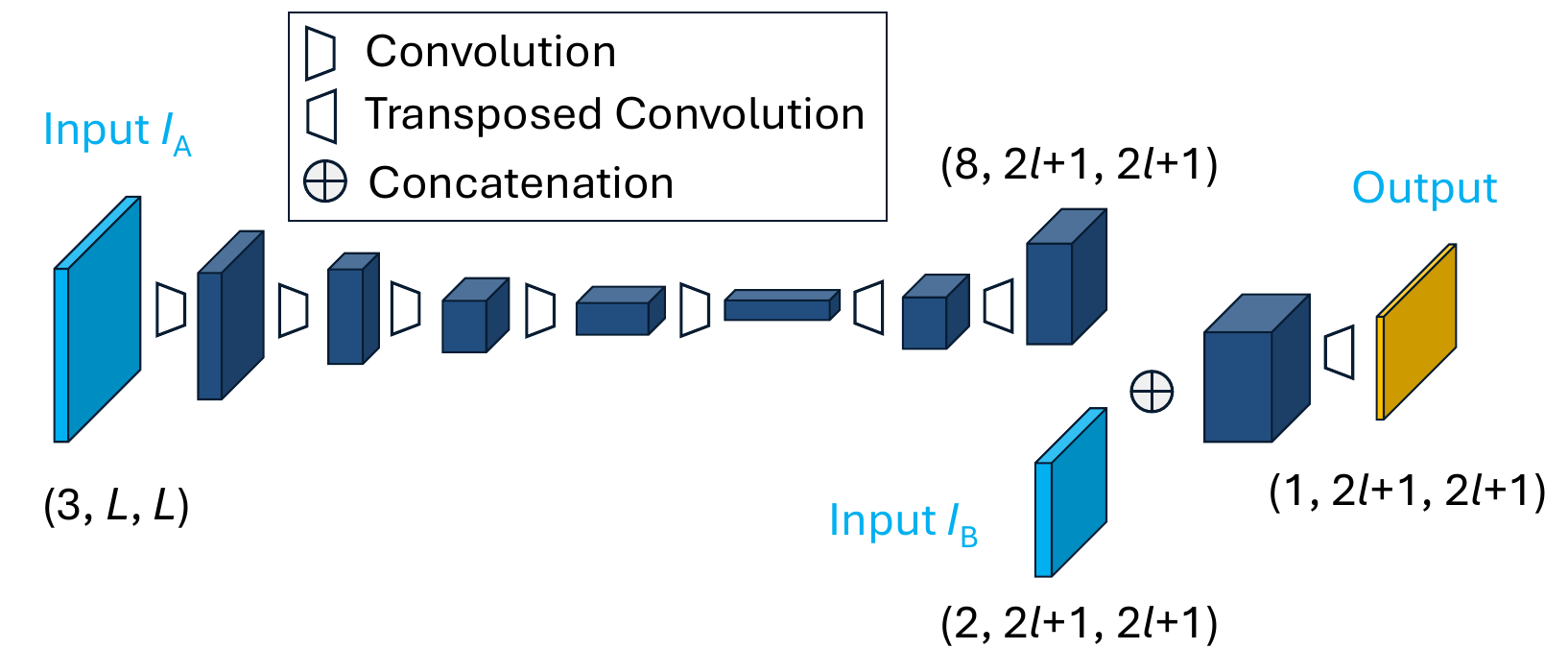} 
    \caption{Proposed Q-network architecture for state-action value prediction. The input is derived from the building height map, the UAV's current location, and the measured RSRPs. The output represents the Q-values corresponding to each possible action within a \( (2l+1) \times (2l+1) \) action space.}
    \vspace{-0.5cm}
    \label{fig:proposed_model}
\end{figure}

The Q-network architecture is demonstrated in Fig.~\ref{fig:proposed_model}.
The model consists of five convolutional layers and three transposed convolutional layers, and it takes two inputs: \( \mathbf{I}_{\text{A}} \) and \( \mathbf{I}_{\text{B}} \). The input \( \mathbf{I}_{\text{A}} \) is a \( 3 \times L \times L \) matrix, where each of the three \( L \times L \) matrices corresponds to the encoded UAV location \( \mathbf{S}_{\text{enc}}^{\text{UAV}}(k) \), the encoded measurements \( \mathbf{Z}_{\text{enc}}(k) \), and the normalized building height array \( \mathbf{M}_{\text{bld}} \).  The second input, \( \mathbf{I}_{\text{B}} \), is a \( 3 \times (2l+1) \times (2l+1) \) matrix, which is a cropped submatrix of \( \mathbf{I}_{\text{A}} \). It is generated by extracting localized regions of \( \mathbf{Z}_{\text{enc}}(k) \) and \( \mathbf{M}_{\text{bld}} \) centered around the UAV's location, as shown in Fig.~\ref{fig:system_model}(b).  
The input \( \mathbf{I}_{\text{A}} \) provides the neural network with a global perspective of the map over a larger area, while \( \mathbf{I}_{\text{B}} \) focuses on a localized region of size \( (2l+1) \times (2l+1) \), corresponding to both the action space and output size. The output of the network is a \( (2l+1) \times (2l+1) \) matrix, where each value represents the Q-value for moving the UAV to the respective location in the action space.


\subsection{Reward}
The policy Q-network, denoted as \(Q_{\text{policy}}\) with parameters~\( \mathbf{\theta} \), predicts the next UAV location \(\hat{\mathbf{S}}^{\text{UAV}}(k+1)\) as follows:
\begin{equation}
\hat{\mathbf{S}}^{\text{UAV}}(k+1) = \operatorname*{arg\,max}_{\mathbf{a} \in \mathcal{A}(k)} Q_{\text{policy}}(\mathbf{h}(k), \mathbf{a}; \mathbf{\theta}).
\label{eqn:predicted_location}
\end{equation}
The reward function is designed such that if \( \hat{\mathbf{S}}^{\text{UAV}}(k+1) \) is not within a CH, it is mildly penalized. However, if \( \hat{\mathbf{S}}^{\text{UAV}}(k+1) \) lies within a building or if the straight line connecting \( \hat{\mathbf{S}}^{\text{UAV}}(k+1) \) and \( \hat{\mathbf{S}}^{\text{UAV}}(k) \) intersects a building, it is heavily penalized. On the other hand, finding a CH is not penalized, as it terminates the task. Mathematically, the reward function is expressed as:
\begin{equation}
r\big(\mathbf{h}(k), \mathbf{a}\big) =
\begin{cases} 
    0 & \text{if } Z(k+1) < \epsilon_{\text{CH}}, \\
    -1.25 & \text{if } \hat{\mathbf{S}}^{\text{UAV}}(k+1) \notin \mathbf{S}_{\text{prm}}(k), \\
    -0.25 & \text{otherwise}.
\end{cases}
\label{eqn:reward}
\end{equation}
Here, \( Z(k+1) \) is the measured RSRP at \(\hat{\mathbf{S}}^{\text{UAV}}(k+1) \), \( \epsilon_{\text{CH}} \) is the coverage hole threshold, and \( \mathbf{S}_{\text{prm}}(k) \) represents the set of permissible grid locations as defined in~\eqref{eqn:S_prm}.

\subsection{Loss Function}
We denote the parameters of the policy network \( Q_{\text{policy}} \) and the target network \( Q_{\text{target}} \) by \( \mathbf{\theta} \) and \( \mathbf{\theta}^- \), respectively. The loss function in a DDQN minimizes the difference between the Q-value predicted by \(Q_{\text{policy}}\) for the current state and the Q-values predicted by \(Q_{\text{target}}\) for the future states. It is defined as follows:
\begin{equation}
L_1(\mathbf{\theta}) = \mathbb{E}_{\mathbf{h}, \mathbf{a}, r, \mathbf{h}'} \left[ \big( y - Q_{\text{policy}}(\mathbf{h}, \mathbf{a}; \mathbf{\theta}) \big)^2 \right],
\label{eqn:ddqn_loss}
\end{equation}
where \( \mathbf{h}, \mathbf{a}, \mathbf{h'} \) represent the current state as in~\eqref{eqn:current_state}, action taken, the next state, \( r \) is the reward as in~\eqref{eqn:reward}, inputs omitted for simplicity, and \( y \), the target, is defined as:
\begin{equation}
y = r + \gamma Q_{\text{target}}\big(\mathbf{h}', \operatorname*{arg\,max}_{\mathbf{a}' \in \mathcal{A}(k+1)} Q_{\text{policy}}(\mathbf{h}', \mathbf{a}'; \mathbf{\theta}); \mathbf{\theta}^-\big), \label{eq:discounted_reward}
\end{equation}
where \( \gamma \) is the discount factor for future rewards.
The target \( y \) is computed using the target network \( Q_{\text{target}} \), while the action selection relies on the policy network \( Q_{\text{policy}} \). This decoupling of action selection and evaluation reduces the overestimation bias of the policy network.

The goal of RL is to maximize the cumulative reward, which corresponds to completing the task in fewer steps, as each step adds a negative reward when the task remains incomplete and the CH is not detected. However, the sparsity of training examples poses challenges to stable training~\cite{jaderberg2016reinforcement}. To address this issue, auxiliary loss terms are often introduced. 
In our case, the frequency of CHs is much lower than that of covered areas. To improve training stability, we add an auxiliary loss term, the squared reward function, to the standard DDQN loss term. The final loss function is expressed as:
\begin{equation}
L(\mathbf{\theta}) = L_1(\mathbf{\theta}) + \alpha r^2,
\label{eqn:final_loss}
\end{equation}
where \( L_1(\mathbf{\theta}) \) is defined in~\eqref{eqn:ddqn_loss}, \( \alpha \) is a hyperparameter, and \( r \) is the reward as in~\eqref{eqn:reward}.

\subsection{Trajectory Prediction}
The algorithm for updating the UAV's trajectory to locate a CH using the proposed DDQN model is outlined in Algorithm~\ref{algo:uav_traj}.
\begin{algorithm}[t]
\caption{UAV Trajectory Update Algorithm}
\begin{algorithmic}[0]  
\State \textbf{Input:} \( \mathbf{S}^{\text{UAV}}(0) \), \( \mathbf{S}_{\text{prm}}(0) \), \(\mathbf{h}(0)\), \(\mathcal{A}(0)\), \(\mathbf{\theta}, k \)
\For{$i = 1, 2, \dots, k$}
    \State \textbf{if} \(Z(i-1) \leq \epsilon_{\text{CH}}\) \textbf{then} \textbf{break}
    \State Obtain \(\hat{\mathbf{S}}^{\text{UAV}}(i)\) in~\eqref{eqn:predicted_location}
    \State \(\mathbf{v} = \hat{\mathbf{S}}^{\text{UAV}}(i) - \mathbf{S}^{\text{UAV}}(i-1)\)
    \State \( \hat{u} = \max_{u \in (0,1)} \left\{ u \mid \mathbf{S}^{\text{UAV}}(i-1) + u \mathbf{v} \in \mathbf{S}_{\text{prm}}(i-1) \right\} \)
    \State \(\mathbf{S}^{\text{UAV}}(i) = \mathbf{S}^{\text{UAV}}(i-1) + \hat{u} \mathbf{v}\)
    \State Compute \(\mathbf{S}_{\text{prm}}(i)\), \( \mathbf{h}(i) \), \( \mathcal{A}(i) \) in \eqref{eqn:S_prm},\,\eqref{eqn:encode_loc}-\eqref{eqn:Ak_eqn}
\EndFor
\end{algorithmic}
\label{algo:uav_traj}
\end{algorithm}
The algorithm utilizes the DDQN model to update the UAV's location for a maximum of \(k\) steps unless a CH is detected before reaching \(k\) steps. For each intermediate waypoint, the DDQN model predicts the UAV's next location, \(\hat{\mathbf{S}}^{\text{UAV}}(i)\). However, if the predicted waypoint intersects with a building obstacle, the UAV is redirected to a valid point along the predicted path, avoiding the obstruction. This adjusted point, denoted as \(\mathbf{S}^{\text{UAV}}(i)\), becomes the new waypoint. The UAV then moves to this point, measures RSRP, and computes state vectors for the next step.

\section{Baseline Non-ML Approaches}
\label{sec:baseline}
The locations of coverage holes in an urban area are largely dependent on the building map, as demonstrated in Fig.~\ref{fig:example}. To compare with the proposed UAV-assisted RL-based method, we consider several baseline predictors derived directly from the building maps described below.
\begin{figure}[t!]
    \centering
    \includegraphics[width=\linewidth,trim={1.2cm 3.7cm 2.0cm 1.4cm},clip]{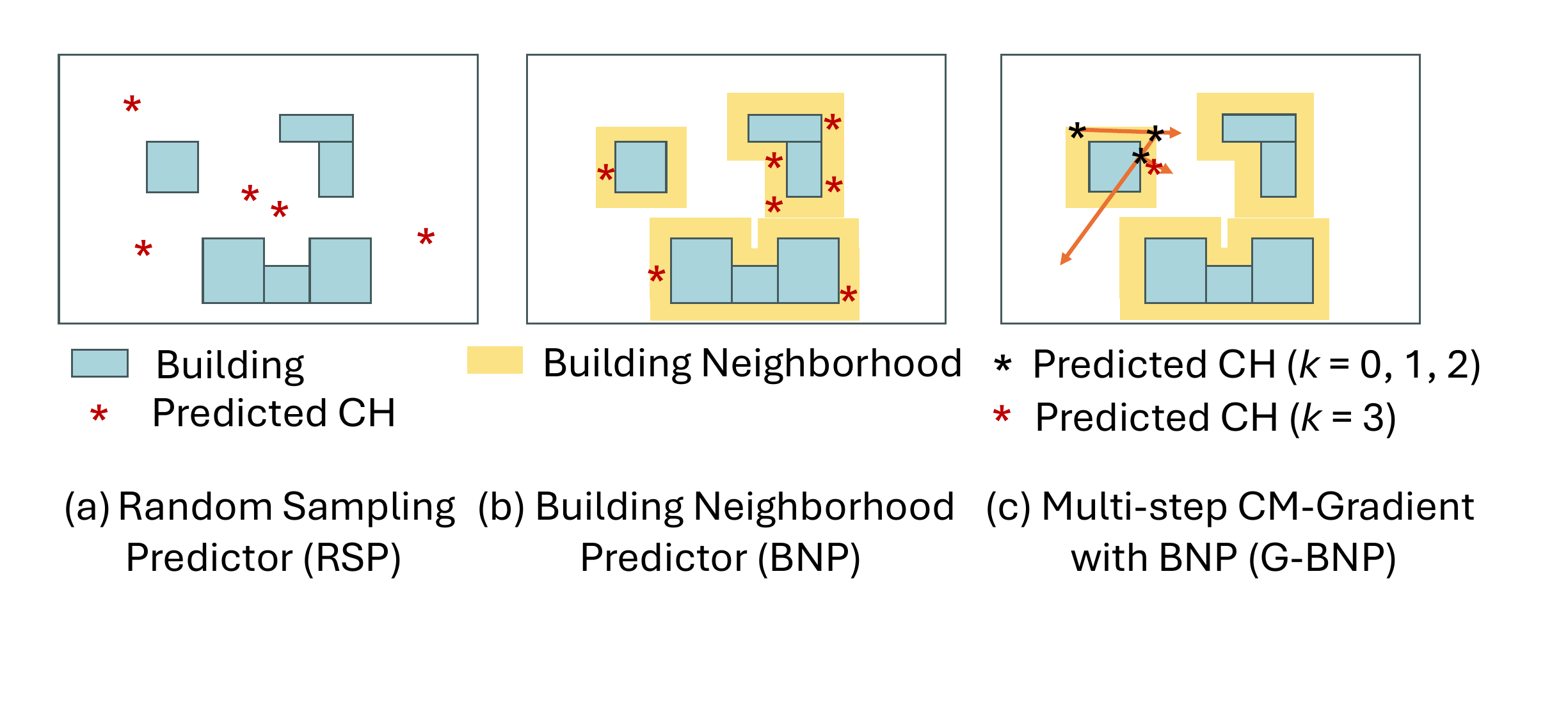} 
    \caption{Non-ML CH predictors derived from the building map: (a) randomly predicts a location outside the buildings, (b) randomly predicts a location within close proximity to a building, and (c) starts from a randomly sampled point within the building neighborhood (\( k=0 \)) and uses the CM gradients to iteratively move in the direction of the negative gradient while remaining bounded within the building neighborhood.}
    \label{fig:non_ml_baseline}
    \vspace{-0.5cm}
\end{figure}
\subsubsection{Random Sampling Predictor (RSP)} 
This approach randomly obtains samples from locations not within the buildings and returns them as the predicted CH, as shown in Fig.~\ref{fig:non_ml_baseline}(a). This serves as a baseline for comparison with other methods.

\subsubsection{Building Neighborhood Predictor (BNP)}
This is another CH predictor that randomly samples only within a given neighborhood of outdoor locations, as shown in the yellow-shaded areas in Fig.~\ref{fig:non_ml_baseline}(b). A hyperparameter \(d_{\text{B}}\) defines the maximum distance from the building walls. We conduct experiments with three settings: \(d_{\text{B}} = \{8 \, \text{m}, 16 \, \text{m}, 32 \, \text{m}\}\).

\subsubsection{Multi-step CM-Gradient Predictor (G-RSP)}
This method starts from an initial location provided by the RSP and uses the CM to follow the gradient descent algorithm to predict the next waypoint. The update rate, a hyperparameter in the gradient descent equation, is set to \(1\), and the location is updated as follows:
\begin{equation}
\mathbf{\hat{S}}_{\text{G-RSP}}^{\text{UAV}}(k+1) = \mathbf{S}^{\text{UAV}}(k) - \frac{\partial Z(k)}{\partial \mathbf{S}^{\text{UAV}}(k)}.
\label{eqn:grsp_update}
\end{equation}
To ensure a fair comparison, the prediction adjustment and iteration steps presented in Algorithm~\ref{algo:uav_traj} are applied, replacing~\eqref{eqn:predicted_location} with~\eqref{eqn:grsp_update} within the algorithm.

\subsubsection{Multi-step CM-Gradient with BNP (G-BNP)}
This method begins at an initial location provided by the BNP and updates the location by following the gradient descent of the CM, similar to the previous method. If the predicted location gets outside of the building neighborhood, it is adjusted within the neighborhood \(d_{\text{nbr}}\), as shown in Fig.~\ref{fig:non_ml_baseline}(c).

The two gradient-based methods are implemented using the computed CM from NVIDIA Sionna. However, these methods are unlikely to mimic real-world scenarios, as computing the gradient through delta movements of the UAV is challenging in the presence of noise. Therefore, the gradient-based methods are presented as theoretical upper bounds, assuming UAVs could measure the gradient accurately in all directions. In practice, the proposed RL method measures only a single RSRP at a waypoint.

\section{Numerical Results}
\label{sec:num_results}
\subsection{Simulation Environment}
We selected a total of \( 13,200 \) locations covering the city of San Jose, California, with each neighboring location spaced \( 512 \, \text{m} \) apart. After downloading the maps from OpenStreetMap, we filtered \( 8,017 \) maps with urban characteristics. 
The downloaded maps had dimensions of \( 1,080 \, \text{m} \times 1,080 \, \text{m} \), with each pixel representing \( 1 \, \text{m} \). We then processed the maps using Blender to prepare them for NVIDIA Sionna Raytracing software. A BS equipped with an omnidirectional antenna was deployed randomly at an altitude of \(2\,\text{m}\) above the ground or building (where available) to compute the RSRP at a fixed height of \(2\,\text{m}\)~\cite{li2024geo2sigmap}. The computed RSRP map provides the ground-truth CM for determining actual CHs, and UAV-measured RSRPs are also simulated from it.

To speed up the convolutional processing time with smaller-sized images, the \( 1,080 \, \text{m} \times 1,080 \, \text{m} \) building maps and coverage maps were center-cropped to \( 484 \, \text{m} \times 484 \, \text{m} \) and subsequently downsampled by a factor of 4, resulting in a \( 121 \times 121 \) grid size.
Using the notations introduced in Section~\ref{sec:sys_model}, this corresponds to an \( L \times L \) grid where \( L = 121 \), and \( H = 2 \, \text{m} \) as the RSRP is computed at a height of \( 2 \, \text{m} \). For other simulation parameters, 
We set \( \epsilon_{\text{CH}} \) to \( -100 \, \text{dB} \), and \( l = 15 \), resulting in a \( 31 \times 31 \) action space for UAV movement per step, representing an area of \( 124 \, \text{m} \times 124 \, \text{m} \), given a pixel resolution of \( 4 \, \text{m} \). Additionally, we set the value of 
\( c \) to \( 0.1 \) in \eqref{eqn:encode_loc} and \eqref{eqn:encode_rsrp}, \( \gamma \) to \( 0.99 \) in \eqref{eq:discounted_reward}, and \( \alpha \) to \( 1.0 \) in \eqref{eqn:final_loss}.

\subsection{Evaluation Metric}
In our simulation environment, a building map contains \( L \times L \) or \( 121 \times 121 \) points, with each point representing a \( 4 \, \mathrm{m} \times 4 \, \mathrm{m} \) grid. The set of CH points, where the RSRP measurement falls below \( \epsilon_\text{CH} \), can be mathematically represented as follows:
\begin{equation}
\mathbf{S}_\text{cmb}^\text{CH} = \{\mathbf{S}(i, j) \in (\mathbb{N}, \mathbb{N}) \mid Z(\mathbf{S}(i, j)) < \epsilon_\text{CH} \},
\end{equation}
where \( \mathbf{S}(i, j) \) represents a grid point, \( i, j \in \{1, 2, \ldots, L\}\), and \(Z(\cdot)\) is the RSRP measurement at a point.
We have four baseline methods presented in Section~\ref{sec:baseline} and our proposed RL-based method.
The proposed RL-based method and each baseline method start with \( N_\text{sam} \) randomly sampled unique points for each map. The \( N_\text{sam} \) initial points result in \( N_\text{sam} \) predicted CHs. Mathematically, the array of predicted CHs can be expressed as a combination of individual CH predictions from each starting point as follows:
\begin{equation}
\mathbf{\hat{S}}_\text{cmb}^\text{CH} = \{\mathbf{\hat{S}}^\text{CH}(i) \in (\mathbb{N}, \mathbb{N}) \mid i \in \{1, 2, \ldots, N_\text{sam}\}\},
\end{equation}
where \( \mathbf{\hat{S}}_\text{cmb}^\text{CH} \) may contain duplicate points, as different starting points near a CH may find the same CH. A subarray of \( \mathbf{\hat{S}}_\text{cmb}^\text{CH} \) containing only true predictions can be defined as follows:
\begin{equation}
\mathbf{\hat{S}}_\text{tru}^\text{CH} = \{\mathbf{\hat{S}}^\text{CH}(i) \in \mathbf{S}_\text{cmb}^\text{CH} \mid i \in \{1, 2, \ldots, N_\text{sam}\}\},
\end{equation}
which may also contain one CH multiple times.

To evaluate the performance of a method, we use two metrics: precision and recall, which are widely used for detecting rare events like CHs. These metrics are mathematically defined as follows:
\begin{equation}
\text{Precision} = \frac{\text{Count}(\mathbf{\hat{S}}_\text{tru}^\text{CH})}{N_\text{sam}}, \quad 
\text{Recall} = \frac{\text{Count}(\text{Unique}(\mathbf{\hat{S}}_\text{tru}^\text{CH}))}{\text{Count}(\mathbf{S}_\text{cmb}^\text{CH})},
\end{equation}
where \( \text{Count}(\cdot) \) represents the number of elements in the array, and \( \text{Unique}(\cdot) \) returns a subarray by removing repeated elements.

\subsection{Simulation Results}
\begin{figure}[!t]
    \centering
    \begin{subfigure}[t]{0.8\linewidth}
        \centering
        \includegraphics[width=\textwidth, trim={0.7cm 0.0cm 1.7cm 1.0cm}, clip]{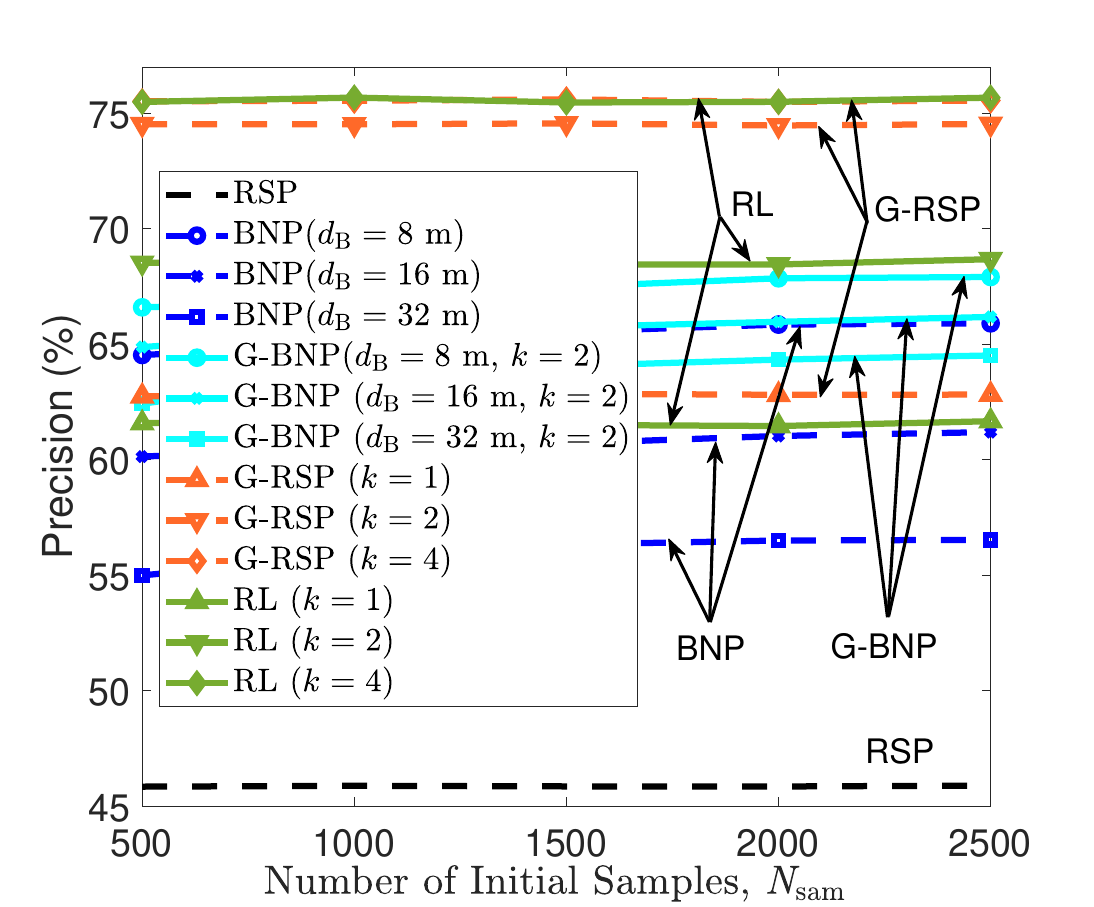} 
        \caption{}
        \label{fig:result_subfig1}
    \end{subfigure}
    \hfill
    \begin{subfigure}[t]{0.8\linewidth}
        \centering
        \includegraphics[width=\textwidth, trim={0.7cm 0.0cm 1.7cm 1.0cm}, clip]{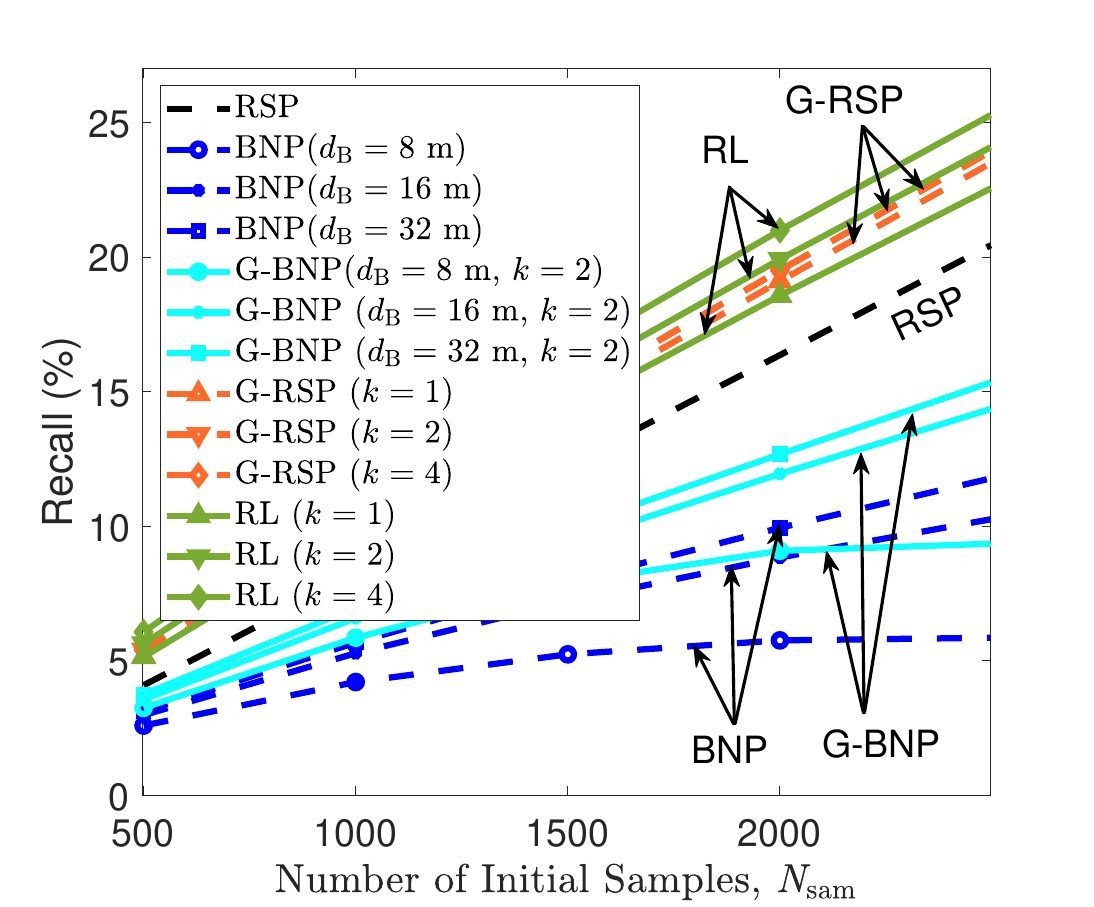} 
        \caption{}
        \label{fig:result_subfig2}
    \end{subfigure}
    \caption{Performance evaluation of the CH detection methods presented in (a) precision and (b) recall.} 
    \label{fig:result}
    \vspace{-0.5cm}
\end{figure}
The precision curves in Fig.~\ref{fig:result}(a) show that BNP significantly improves prediction precision compared to RSP. This indicates that CHs are more likely near buildings, making predictions within an \(8\) m proximity statistically more accurate. As the proximity loosens, precision converges to that of RSP. In contrast, G-RSP achieves very high precision, particularly for \( k > 2 \), where \( k \) represents the number of iterations. Notably, the improvement from \( k = 1 \) to \( k = 2 \) is greater than that from \( k = 2 \) to \( k = 4 \). G-BNP shows lower precision compared to G-RSP for \( k = 2 \). This is because confining the predictions within the building neighborhood, as shown in Fig.~\ref{fig:non_ml_baseline}(c), limits the use of gradient information and potentially causes the algorithm to oscillate at boundaries. Finally, the RL-based method performs slightly less effectively compared to G-RSP for \( k = 2 \) but matches precision at \( k = 4 \), demonstrating the power of RL using only \( 4 \) measurements.

The recall plots demonstrated in Fig.~\ref{fig:result}(b) show an increasing trend with the x-axis, indicating the detection of new CHs as the number of predictions increases for all methods. However, the BNP and G-BNP exhibit slower recall growth due to their inability to detect CHs outside the building neighborhood, inherent to their algorithmic constraints. Consequently, these methods plateau after identifying all CHs within the neighborhood, which also explains their lower recall compared to RSP. Despite this, G-BNP achieves better recall than BNP. Finally, the RL-based method outperforms G-RSP in recall for \( k = 2 \) and \( k = 4 \). Unlike G-RSP, for which the recall saturates between \( k = 2 \) and \( k = 4 \), the RL method continues to improve the recall with increasing \( k \), demonstrating the greater potential for identifying unique CHs.

\section*{Discussion and Conclusion}
This work explores the potential of an RL-powered UAV for autonomous CH detection in urban areas. We proposed an RL method for detecting CHs using building maps and a limited number of RSRP measurements, adaptable to any CH threshold for training. By varying the CH threshold, the approach can detect CHs or identify areas falling below specific signal strength levels to ensure QoS.  Our results demonstrate that the proposed RL-based method is both accurate and efficient, achieving comparable or better precision and recall than the best baseline, G-RSP, while requiring significantly less data. With only four signal measurements, the RL method matches the performance of having coverage map gradients. While the proposed approach can be carried out using a lightweight UAV-based UE, an autonomous UGV-based UE can also be used with potentially different restricted areas. Future work will involve evaluating performance with noisy coverage maps, different building types, and densities, deploying real UAVs, and varying grid resolutions, map size, and action space size.




\bibliographystyle{IEEEtran}
\bibliography{main}

\end{document}